\documentclass[aps,prapplied,reprint,amsfonts,amssymb,amsmath]{revtex4-1}

\usepackage{graphicx}
\usepackage{hyperref}
\hypersetup{
	colorlinks=true, 
	citecolor=blue, 
	urlcolor=blue, 
	linkcolor=blue
}

\newcommand{\eq} [1] {Eq.~(\ref{#1})}
\newcommand{\fig}[1] {Fig.~\ref{#1}}
\newcommand{\tab}[1] {Table~\ref{#1}}
\newcommand{\app}[1] {Appendix~\ref{#1}}

\begin{document}
\title{Graphene-Semiconductor Contact}
\author{M. Javadi}
\email[]{m\_javadi\_b@ut.ac.ir}
\affiliation{Department of Physics, University of Tehran, Tehran 14395-547, Iran}
\date{\today}
\begin{abstract}
A systematic treatment of graphene-semiconductor junction is presented. Finite density of states at the Fermi level of graphene leads to exotic electronic properties at graphene-semiconductor interface. Quite generally, the Schottky-Mott limit and the sum rule of barrier heights are violated due to the internal potential of graphene. By merging the principal characteristics of semiconductor-semiconductor and metal-semiconductor junctions, the graphene-semiconductor contact may be considered as an archetype of unprecedented \emph{semimetal-semiconductor junction}. Generalized interface equations disclose the coupling of junction characteristics with the density of charge carriers in graphene. It will be shown that the well-known effect of tunable barrier height is directly related to the junction capacitance. Furthermore, the relative impact of image-force effect in the presence of barrier tunability is investigated. Experimental methods to gain built-in potential, barrier height, and capacitance of the junction are discussed in detail. Also the development of strong inversion layer at the interface of graphene-silicon samples is uncovered by analyzing experimental data.  
\end{abstract}
\maketitle 
\section{Introduction}
\label{sec:intro} 
Integrating graphene and related two-dimensional (2D) materials into silicon technology is expected to address a considerable share in the next generation of electronic and optoelectronic devices \cite{akinwande_2019}. Regarding the demand for nano-size dimensional scaling, two-dimensional materials with outstanding structural and electronic potentials promise new device concepts with upgraded performance beyond those running the existing technology. Incorporating graphene into complementary metal oxide semiconductor (CMOS) technology \cite{Kim_2011} and innovative graphene/semiconductor hybrid devices such as gate tunable Schottky barrier diodes \cite{Yang_2012, LaGasse_2019}, as well as graphene/silicon rectifiers \cite{Di_Bartolomeo_2016, Chen_2011, Sinha_2014}, optical detectors \cite{Amirmazlaghani_2013, An_2013, Liu_2014, Trushin_2018, Goykhman_2016}, and solar cells \cite{Behura_2019, Li_2010, Miao_2012, Won_2018} may be considered as the preliminary building blocks of upcoming semiconductor technology. The integration of 2D and 3D materials also release new physical concepts on the nature of contact \cite{Xu_2016} and charge carrier transfer at 2D/3D interface \cite{Liang_2015, Ang_2016, Ang_2017, Ang_2018, Ang_2019, Zho_2018, Tsai_2019, Liu_2019}. Likewise, bias modulation of barrier height \cite{Tongay_2012,An_2013, Di_Bartolomeo_2_2016} and self-gating effect \cite{Di_Bartolomeo_2017} at graphene-semiconductor contact. 

As an essential ingredient of 2D-3D integration, understanding the interface physics at graphene-semiconductor contact is of great importance. In addition, accurate description of junction properties is vital for develop new device concepts based on graphene-semiconductor junction. From a general perspective, this junction may be handled in a way akin to the standard model of metal-semiconductor contact. However, finite density of states in the vicinity of Dirac point makes graphene Fermi level very sensitive upon close proximity to other semiconductors. This in turn leads to the emergence of exotic electronic properties at the interface.

The fundamental dissimilarity between metal-semiconductor and graphene-semiconductor contacts can be understood as follows. When a metal makes contact with a semiconductor, an ohmic or rectifying (Schottky) junction may be developed at the interface. The necessary condition for the formation of Schottky junction on a n-type (p-type) semiconductor is $q\phi_m>q\phi_s$ ($q\phi_m<q\phi_s$) where $q\phi_m$ and $q\phi_s$ are the work functions of metal and semiconductor, respectively ($q$ is the elementary charge). Under an ideal condition (Schottky-Mott limit), the barrier height for charge injection from metal into n- and p-type semiconductors is given by $q\varphi_{Bn}=q(\phi_m-\chi)$ and $q\varphi_{Bp}=E_{gap}-q(\phi_m-\chi)$ where $q\chi$ and $E_{gap}$ are electron affinity and band gap energy of the semiconductor \cite{Sze_2006}. In thermal equilibrium, a built-in potential of $\psi_{bi}=\left|\phi_m-\phi_s\right|$ is developed across the junction. The variation of metal Fermi level upon junction formation (due to charge exchange with semiconductor) is proportional to $\delta E_F/E_F\propto N/n$ \cite{Di_Bartolomeo_2016}, where $N\!\sim\! 10^{13}-10^{18}\ cm^{-3}$ is the concentration of dopants in semiconductor. Because of the high density of charge carriers in metal ($n\!\sim\!10^{22}\ cm^{-3}$), the Fermi level does not vary, and hence, the entire built-in potential is established at the semiconductor surface.\\
A major difference between graphene (as a semimetal) and a normal metal is that the density of states at Fermi level of the former is vanishingly small. Owing to this property, the process of charge exchange upon contact formation with a semiconductor would alter the Fermi level of graphene. In other words, the Fermi level in graphene-semiconductor system is different than the Fermi level of isolated graphene. Likewise, when an external bias voltage is applied to the graphene-semiconductor system, the graphene Fermi level shifts with respect to its equilibrium position. This effect causes modifications in the built-in potential distribution, barrier height, and capacitance of junction.\\
\begin{figure*}[]
	\includegraphics[scale=1]{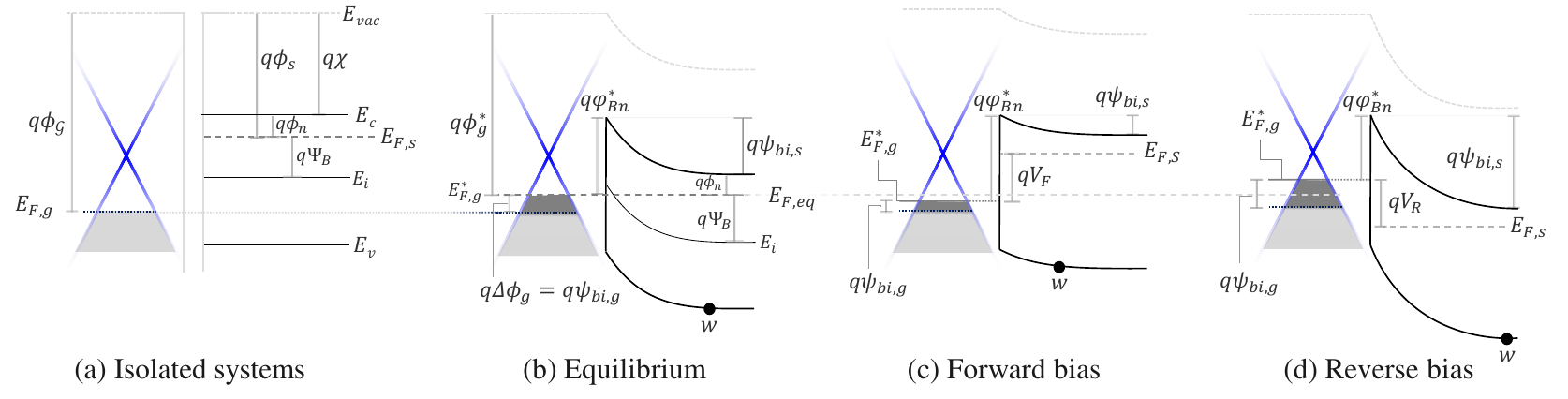}
	\caption{
		\label{fig:bandalign}
		Energy-band diagrams at graphene$-$n-type semiconductor interface: (a) graphene and semiconductor as separate (isolated) systems, (b) connected into a unit system at thermal equilibrium, (c) at forward bias, and (d) at reverse bias. Various parameters are as follows. $E_{vac}$: vacuum level; $q\phi_{g}$ and $E_{F,g}$: work function and Fermi level of isolated graphene; $q\phi_s\text{,}\ q\chi\text{,}\ E_{i}\text{,}\ E_{F,s}\text{,}\ E_C\text{,}\ E_V\text{, and}\ E_{gap}$: work function, electron affinity, intrinsic Fermi level, (extrinsic) Fermi level, conduction band edge energy, valance band edge energy, and band gap of the semiconductor, respectively. 
	}
\end{figure*}
This study provides a systematic assessment of junction properties at ideal graphene-semiconductor contact. Basic energy relations as well as major electronic characteristics are presented and differences with the conventional metal-semiconductor contact are addressed. Junction properties in the various limits of undoped, doped, and highly-doped graphene are also investigated. For the sake of practical illustrations, the model is applied to graphene-silicon contact. The relative influence of image-force effect on the barrier tunability as well as various phases of depletion, weak inversion, and strong inversion at graphene-silicon interface are presented. 
\section{Energy relations}
\label{sec:energy}
Energy-band diagrams of an isolated graphene and a n-type semiconductor are shown in \fig{fig:bandalign}(a). We will consider non-ohmic contact where the work function of isolated graphene ($q\phi_{g}$) is higher than that of the semiconductor. When the two systems are brought together so that the gap between them is small enough to be transparent to charge carriers, electrons will flow from the semiconductor into graphene in order to line up Fermi level through the whole system. This charge flow is finally quenched by an internal (built-in) potential developed at the interface. Because of the finite density of states in the vicinity of Dirac point, this charge transfer would alter the Fermi level of graphene from its previous position in the isolated system. Indeed, when graphene makes contact with a n-type semiconductor, its work function \textit{decreases} relative to the work function of the isolated graphene. Assuming a net change of $q\Delta\phi_{g}$, from the energy-band diagram shown in \fig{fig:bandalign}(b)
\begin{equation}
	\label{eq:g-n:gshift}
	q\phi_{g}^{*}=q(\phi_{g}-\Delta\phi_{g})\;,
\end{equation}
where $q\phi_{g}^{*}$ is the work function of graphene in the graphene-semiconductor system (superscript asterisk is used to clarify parameters related to the graphene-semiconductor system in the whole manuscript). The barrier height at the interface is given by
\begin{equation}
	\label{eq:g-n:schottky}
	q\varphi_{Bn}^{*}=q(\phi_{g}^{*}-\chi)=q(\phi_{g}-\chi) - q\Delta\phi_{g}\;.
\end{equation} 
Alternatively, the barrier height is also given by
\begin{equation}
	\label{eq:g-n:schottky2}
	q\varphi_{Bn}^{*}=q(\psi_{bi,s}+\phi_{n})\;,
\end{equation}
where $\psi_{bi,s}$ is built-in potential developed at the semiconductor surface and $q\phi_n=q(\phi_s-\chi)$. Eliminating $q\varphi_{Bn}^*$ from \eq{eq:g-n:schottky} and \eq{eq:g-n:schottky2}, the total built-in potential is obtained as
\begin{equation}
	\label{eq:g-n:builtin}
	\psi_{bi}\equiv\phi_{g}-\phi_s = \psi_{bi,s} + \Delta\phi_{g}\;.
\end{equation}
In this regard, the net change in the work function of graphene can be considered as a part of the total built-in potential that is dropped across graphene layer, i.e. $\Delta\phi_{g}=\psi_{bi,g}$. \eq{eq:g-n:schottky} highlights one of the main outcomes of finite density of states in graphene: the equilibrium barrier height at graphene-semiconductor interface is generally smaller, by an amount of $q\psi_{bi,g}$, when compared with a normal metal-semiconductor junction $q\varphi_{Bn}^{*}=q\varphi_{Bn}-q\psi_{bi,g}$.

Similarly, when graphene makes non-ohmic contact with a p-type semiconductor ($q\phi_{g}<q\phi_s$), the equilibrium condition of Fermi level alignment is fulfilled by the transfer of positive carriers (holes) from semiconductor into graphene. In this case, the work function of graphene in the graphene-semiconductor system \textit{increases} relative to the work function of isolated graphene. In the same manner of n-type semiconductor, the electronic energy relations for graphene$-$p-type semiconductor junction may be obtained as
\begin{subequations}
	\label{eq:g-p}		
	\begin{eqnarray}
		\label{eq:g-p:gshift} 
		q\phi_{g}^{*}\ && =q\phi_{g}+q\psi_{bi,g}\;, \\
		\label{eq:g-p:schottky}
		q\varphi_{Bp}^{*}\ && =E_{gap}-q(\phi_{g}-\chi)-q\psi_{bi,g}\;,\\
		\label{eq:g-p:builtin}
		\psi_{bi}\ && =\phi_s-\phi_{g}=\psi_{bi,s}+\psi_{bi,g}\;.
	\end{eqnarray}
\end{subequations}
As another basic difference, it is worth to note that for the metal-semiconductor junctions the sum of barrier heights on n-type and p-type semiconductors is equal to the band gap of semiconductor $q\varphi_{Bn}+q\varphi_{Bp}=E_{gap}$. This sum rule is normally robust against non-ideal factors such as interface states \cite{Sze_2006, Rhoderick_1982}. In comparison, at graphene-semiconductor interface by combining \eq{eq:g-n:schottky} and \eq{eq:g-p:schottky} we obtain
\begin{equation}
	\label{eq:schottky_total}
	q\varphi_{Bn}^{*}+q\varphi_{Bp}^{*}=E_{gap} - 2q\psi_{bi,g}\;.
\end{equation}
Hence, the sum rule of barrier heights on n- and p-type substrates is smaller than the band gap energy (by an amount of $2q\psi_{bi,g}$). This conclusion is observed experimentally, for example see Ref. \onlinecite{Chen_2011}.
\section{Built-in potential}
\label{sec:builtin}
\begin{figure}[]
	\includegraphics[scale=1.1]{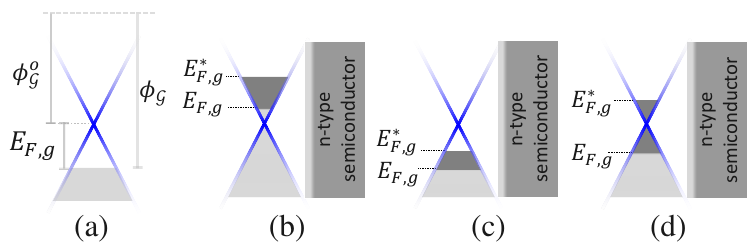}
	\caption{
			\label{fig:GWorkfunction}
			(a) Work function and Fermi level of isolated graphene. (b-d) Fermi level shift upon junction formation with a n-type semiconductor when the isolated graphene is (b) n-doped, (c) p-doped and the contact formation does not affect doping polarity of graphene, and (d) same as previous but the transferred charge changes the polarity of graphene.
			}
\end{figure}
From energy perspective, the built-in potential of graphene is given by the Fermi level shift (\fig{fig:bandalign}(b))
\begin{equation}
	\label{eq:GBuiltin}
	q\psi_{bi,g}=\left|\Delta E_{F,g}\right|=\left| E_{F,g}^*-E_{F,g} \right|\;,
\end{equation}
where $E_{F,g}$ is the Fermi level of isolated graphene. It is convenient to define Fermi level with respect to the band-crossing point. As schematically shown in \fig{fig:GWorkfunction}(a), the work function of graphene may be written as
\begin{equation}
	\label{eq:GWorkfunctionFermilevel}
	q\phi_{g}=q\phi_{g}^o + p_g\left| E_{F,g} \right|\;,
\end{equation}
where $q\phi_{g}^o$ is the work function of neutrality point and $p_g=\pm1$ shows the doping polarity of graphene (+1 for p-doped and -1 for n-doped). The same convention will also be used to clarify the doping of semiconductor ($p_s=\pm1$).

Within the framework of Dirac cone approximation, the Fermi energy may be written as $E_{F,g}=\hbar v_F \sqrt{\pi n}$ \cite{Castro_Neto_2009, Das_Sarma_2011}, where $v_F$ and $n$ are the Fermi velocity and charge carrier density of isolated graphene, respectively. Upon junction formation, a charge density of $\Delta n=Nw$ will be transferred form the semiconductor into graphene. Here, $N$ and $w$ are the concentration of dopants and the width of depletion region at the semiconductor surface. \fig{fig:GWorkfunction}(b-d) shows all possible scenarios for the shift of Fermi level upon contact formation with a n-type semiconductor. For an isolated n-doped graphene (\fig{fig:GWorkfunction}(b)), the built-in potential is given by ($p_{g}=-1\text{, } p_s=-1$)
\begin{equation}
	\label{eq:FermiPg-1Ps-1}
	q\psi_{bi,g}=\hbar v_F\sqrt{\pi (n + Nw)}-\hbar v_F \sqrt{\pi n}\;.
\end{equation}
The polarity of p-doped graphene may be altered upon contact formation with a n-type semiconductor. The final polarity critically depends on the ratio of $n/(Nw)$. For $n/(Nw)\!>\!1$ the polarity of graphene in the graphene-semiconductor system would be the same as of the  isolated graphene ($p_g^*=p_g$, \fig{fig:GWorkfunction}(c)). On the other hand, for $n/(Nw)\!<\!1$ the polarity would flip as a result of contact formation ($p_g^*=-p_g$, \fig{fig:GWorkfunction}(d)). In this case, the internal potential is given by ($p_{g}=+1\text{, } p_s=-1$)
\begin{eqnarray}
	\label{eq:FermiPg+1Ps-1}	
	q\psi_{bi,g} =\ && \hbar v_F\sqrt{\pi n} -\nonumber\\
	&& sgn(n-Nw)\hbar v_F\sqrt{\pi \left| n - Nw \right|}\;,
\end{eqnarray} 
where $sgn(x)$ is the sign function. Similar scenarios can be written for graphene$-$p-type semiconductor interface. Gathering all the possible scenarios into a single equation, the graphene built-in potential is given by
\begin{equation}
	\label{eq:FermiGeneral}
	q\psi_{bi,g}=p_{gs}\hbar v_F \sqrt{\pi n}\left(\zeta\sqrt{\left|1+p_{gs}\frac{Nw}{n}\right|}-1\right)\;,
\end{equation}
where $p_{gs}=p_g p_s$ and the pre-factor $\zeta$ is defined as 
\begin{equation}
		\zeta= \begin{cases}
		1 &p_{gs}=+1\\
		sgn(n-Nw) &p_{gs}=-1
		\end{cases}\;.
\end{equation} 
\begin{figure*}[]
	\includegraphics[scale=1.0]{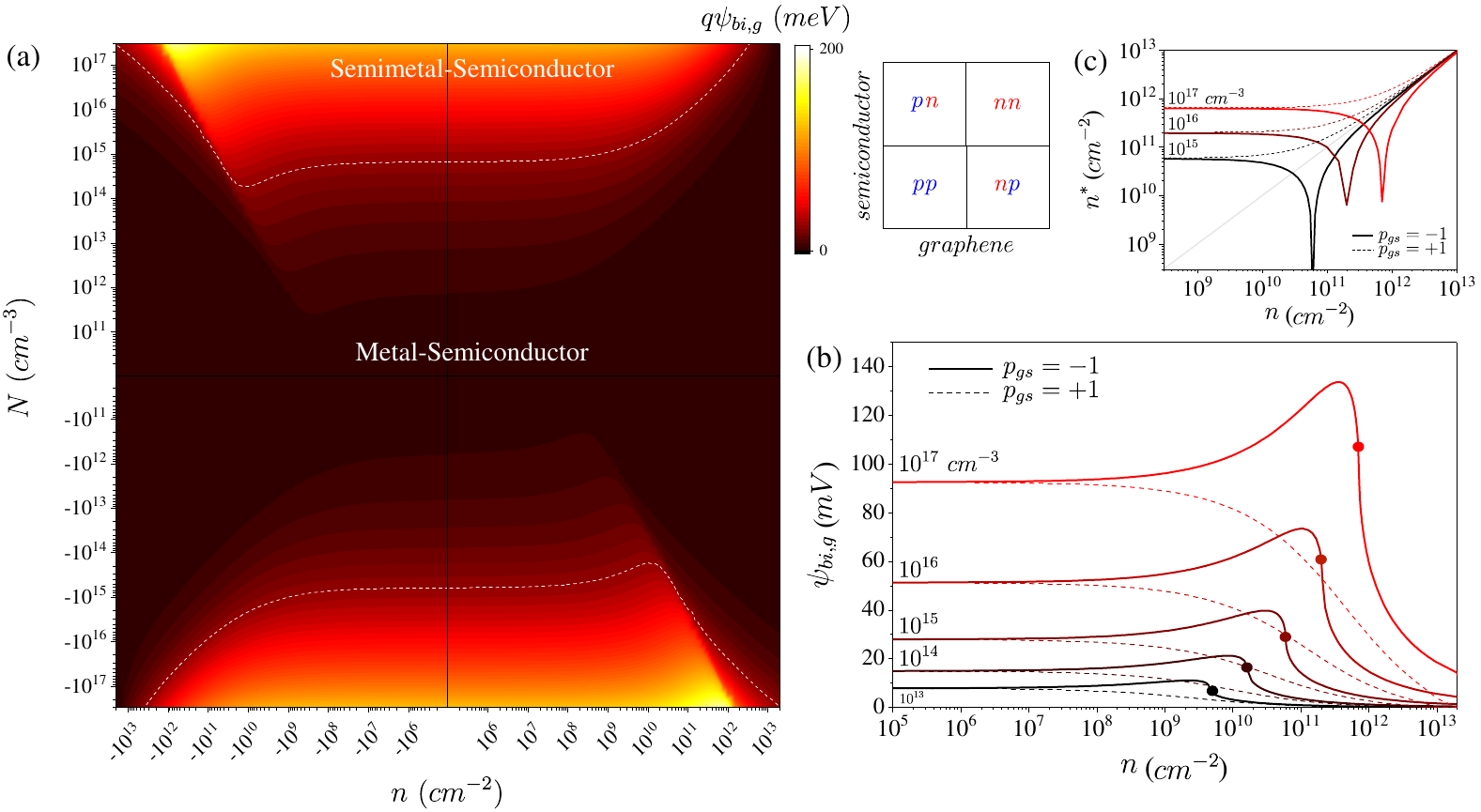}
	\caption{
			\label{fig:MAP} 
			(a) Equilibrium (zero-bias) built-in potential energy of graphene at graphene-semiconductor system. The white dash-lines represent thermal energy level at room temperature ($kT=26\ meV$). Polarity convention is shown in the right upper panel where the first (second) letter represent the polarity of graphene (semiconductor). Electronic properties of crystalline silicon are used for the semiconductor side and the work function of undoped graphene is assumed as $q\phi_{g}^{o}=4.6\ eV$ (see \app{appendixX}). (b) The dependence of $\psi_{bi,g}$ on the polarity of graphene and semiconductor ($p_{gs}$). (c) Density of charge carriers in graphene following contact formation versus the density of carriers in isolated graphene.
			}
\end{figure*}
From \eq{eq:FermiGeneral} it is obvious that the only missing parameter is the width of depletion region ($w$). It is necessary to note that owing to the atomic thickness of graphene as well as its semi-metallic nature, the depletion region is thoroughly developed at the semiconductor surface. Utilizing Poisson equation, the built-in potential of semiconductor is related to the depletion width through 
\begin{equation}
	\label{eq:SBuilt-in}
	q\psi_{bi,s}=\dfrac{q^2 N w^2}{2 \epsilon_s}\;,
\end{equation}
where $\epsilon_s$ is the dielectric constant of semiconductor. Now, the width of depletion region can be obtained by combining Eqs. (\ref{eq:g-n:builtin}), (\ref{eq:g-p:builtin}), (\ref{eq:FermiGeneral}), and (\ref{eq:SBuilt-in})
\begin{eqnarray}
	\label{eq:width}
	q\psi_{bi}\ && =  q\left| \phi_{g} - \phi_s \right| \\
			   && = p_{gs}\hbar v_F \sqrt{\pi n}\left(\zeta\sqrt{\left|1+p_{gs}\frac{Nw}{n}\right|}-1\right) + \dfrac{q^2 N w^2}{2 \epsilon_s}\;.\nonumber
\end{eqnarray}
Note that the work function of semiconductor ($q\phi_s$) is determined from the concentration of dopants (see \app{appendixA}). Hence, the junction properties at graphene-semiconductor interface is thoroughly determined by the density of carriers in the isolated graphene and the concentration of dopants in the semiconductor. \eq{eq:width} may be solved numerically to find the width of depletion region which in turn may be used to determine the built-in potential of graphene [\eq{eq:FermiGeneral}]. \fig{fig:MAP}(a) represents the results of numerical calculations for graphene-silicon contact over the broad range of carrier concentration and various selections of doping polarities. Note that the concentration of holes is shown by negative values. Also note that the experimental accessible values of $n$ is fairly between $n\approx10^{9}$ - $10^{13}\ cm^{-2}$ (intrinsic density of charge carriers at room temperature is $n=9\times10^{10}\ cm^{-2}$ \cite{Fang_2007}). The values beyond this interval are shown for the purpose of symmetry illustration. 

Before to proceed further, it is worth to mention that since built-in potential is developed at the both sides of junction [\eq{eq:g-n:builtin}], form energy perspective the graphene-semiconductor interface is similar to \emph{semiconductor-semiconductor} junction (e.g. a p-n junction). On the other hand, from the depletion region view, since this zone is exclusively developed at the semiconductor side, the graphene-semiconductor interface resembles a Schottky or \emph{metal-semiconductor} junction. In this regard, the graphene-semiconductor contact combines the basic features of the both types of junctions and it is meaningful to refer it as \emph{semimetal-semiconductor} junction. However, we note that
the junction at graphene-semiconductor interface may be classified as semimetal-semiconductor or metal-semiconductor junction depending on $n$ and $N$. The justification parameter is the internal potential of graphene. As far as a non-negligible built-in potential is developed across graphene layer, the contact is identified as semimetal-semiconductor junction. On the other hand, for a negligible $\psi_{bi,g}$ the interface is categorized under metal-semiconductor junction. In \fig{fig:MAP}(a), the magnitude of $q\psi_{bi,g}$ with respect to the thermal energy at room temperature is used to separate the two junction phases at \emph{equilibrium} graphene-silicon interface. For undoped/lightly-doped silicon ($N<10^{14}\ cm^{-3}$), the interface behaves like a normal metal-semiconductor junction. For the higher dopant concentrations ($N>10^{14}\ cm^{-3}$) and $n<10^{13}\ cm^{-2}$, the interface turns into the semimetal-semiconductor junction. As expected, the graphene-silicon contact tends to metal-semiconductor junction with the increment of charge carrier density in the isolated graphene. 

The doping polarity of graphene and semiconductor have a critical role in the magnitude of $\psi_{bi,g}$. For the identical polarities $p_{gs}=+1$, the Fermi level of graphene shifts away from the band-crossing point upon contact formation. In contrast, for $p_{gs}=-1$ the shift direction is toward Dirac point with vanishingly small density of states which results in a quick variation of $E_{F,g}$, and hence, a substantial increment of $\psi_{bi,g}$. In the latter case, there is a critical density of charge carrier in the isolated graphene ($n_{c}$) below which the polarity of graphene would flip upon junction formation, i.e. $p_g^*=-p_g=p_s$ [\fig{fig:GWorkfunction}(d)]. The corresponding points of polarity flip are shown by filled circles in \fig{fig:MAP}(b). In the case of graphene-silicon contact and for the dopant concentrations of $N\!=\!10^{15}$, $10^{16}$, and $10^{17}\ cm^{-3}$, the critical densities are obtained as $n_c\!=\!6\times\!10^{10}$, $2\times\!10^{11}$, and $7\times10^{11}\ cm^{-2}$, respectively (\fig{fig:MAP}(c)).

We note that for specific $p_g$ and $n$, the internal potential of graphene may be different on n-type and p-type semiconductors (see \eq{eq:FermiGeneral} and \fig{fig:MAP}(b)). In this regard, the sum rule of barrier heights [\eq{eq:schottky_total}] may be written in a more general form as
\begin{equation}
	\label{eq:schottky_total2}
	q\varphi_{Bn}^{*}+q\varphi_{Bp}^{*}=E_{gap} - q(\psi_{bi,g}^n+\psi_{bi,g}^p)\;,
\end{equation}
where $\psi_{bi,g}^n$ and $\psi_{bi,g}^p$ denote the built-in potential of graphene on n-type and p-type substrates. \eq{eq:schottky_total2} reduces to \eq{eq:schottky_total} for undoped graphene (see \fig{fig:MAP}(b)).\\
It is also worth to mentioned that the work function of undoped graphene matches very well with the intrinsic Fermi level of silicon (the middle of band gap) which is the reason for symmetrical distribution of $q\psi_{bi,g}$ shown in \fig{fig:MAP}(a). The distribution is less symmetric and asymmetric at graphene-GaAs ($q\chi=4.07\ eV$, $E_{gap}=1.42\ eV$) and graphene-Ge ($q\chi=4.00\ eV$, $E_{gap}=0.67\ eV$) contacts, respectively.  
\section{Two special limits}
\label{sec:twolim}
It is constructive to consider two extreme limits, namely the case of undoped graphene ($n\!\ll\!Nw$), and the other limit of doped graphene where the density of charge carriers in the isolated graphene is much higher than the density of transferred charge due to junction formation ($n\!\gg\!Nw$). Let us to consider the limit of doped graphene first. This case is often the dominant situation in most practical samples where a chemical vapor deposited (CVD-grown) graphene is transferred onto the semiconductor surface by wet-methods \cite{Lemaitre_2012}. The validation domain of $n\!\gg\!Nw$ is justified in \app{appendixB}. Doped graphene is characterized by $\zeta=1$ and $p_g^*=p_g$. In this case, \eq{eq:FermiGeneral} can be approximated as
\begin{equation}
	\label{eq:FermiHighdoped}
	q\psi_{bi,g} = \frac{q^2}{\epsilon_g} \lambda_g Nw\;,
\end{equation}
where $\epsilon_g$ is the dielectric constant of graphene and $\lambda_g$ is the Thomas-Fermi length of charge carriers in isolated graphene 
\begin{equation}
	\label{eq:TomasFermi}
	\lambda_{g}= \frac{\epsilon_g}{q^2} \frac{\partial \mu}{\partial n} = \frac{\epsilon_g}{2q^2}\hbar v_F \sqrt{\frac{\pi}{n}}\;,
\end{equation}
with $\mu \equiv E_{F,g}$. \eq{eq:width} reduces to a quadratic equation whose solution is given by  
\begin{equation}
	\label{eq:width:doped}
	w = \sqrt{(\frac{\epsilon_s}{\epsilon_g}\lambda_{g})^2+\dfrac{2 \epsilon_s \psi_{bi}}{e N}}-\frac{\epsilon_s}{\epsilon_g}\lambda_{g}\;.
\end{equation}
This equation shows that depletion width at the graphene-semiconductor contact is coupled to the screening length of charge carriers in graphene. We note that \eq{eq:width:doped} is indeed a generalization of depletion width at metal-semiconductor junction. At the limit of highly-doped graphene, Thomas-Fermi length vanishes $\lambda_{g}\rightarrow 0$, and the depletion width is given by the conventional equation of $w=\sqrt{2\epsilon_s\psi_{bi}/(qN)}$.

For the case of undoped graphene ($n\ll Nw \text{, } p_g= \text{undefined}$), \eq{eq:FermiGeneral} reduces to
\begin{equation}
	\label{eq:FermiUndoped}
	q\psi_{bi,g} = \hbar v_F \sqrt{\pi Nw}\;.
\end{equation}
In this limit, the polarity of graphene after contact formation is determined by the doping polarity of the semiconductor ($p_g^*=p_s$). \eq{eq:width} for the width of depletion region may be written as 
\begin{equation}
	\label{eq:width:undoped}
	w^2 + c_1 \sqrt{w} - c_2 = 0\;,
\end{equation} 
where $c_1$ and $c_2$ are positive constants given by
\begin{subequations}
	\label{eq:undopedg}		
	\begin{eqnarray}
	\label{eq:undopedg:c1} 
	c_1\ && =\frac{2\epsilon_s\hbar v_F}{q^2}\sqrt{\frac{\pi}{N}} \;, \\
	\label{eq:undopedg:c2}
	c_2\ && =\frac{2\epsilon_s\psi_{bi}}{qN} \;.
	\end{eqnarray}
\end{subequations}
\eq{eq:width:undoped} can be solved numerically or by graphical intersection of $x^4$ and $c_2-c_1 x$ curves ($x=\sqrt{w}$). However, we note that following contact formation the Fermi level of graphene is given by $E_{F,g}^*=\hbar v_F \sqrt{\pi n^*}$ with $n^*=Nw$. By introducing extrinsic Thomas-Fermi length as
\begin{equation}
	\label{eq:exTomasFermi}
	\lambda_{g}^{*}=\frac{\epsilon_g}{2 q^2}\hbar v_F \sqrt{\frac{\pi}{Nw}}\;,
\end{equation}
the built-in potential of graphene [\eq{eq:FermiUndoped}] may be rewritten in the form
\begin{equation}
	\label{eq:FermiUndoped2}
	q\psi_{bi,g}=2\frac{q^2}{\epsilon_g}\lambda_{g}^{*}Nw\;.
\end{equation}
Accordingly, \eq{eq:width:undoped} can be converted into
\begin{equation}
	\label{eq:width:undoped2}
	w^2+4(\frac{\epsilon_s}{\epsilon_g}\lambda_{g}^{*})w-\frac{2\epsilon_s\psi_{bi}}{qN}=0\;,
\end{equation}
which leads to 
\begin{equation}
	\label{eq:undoped2}
	w=\sqrt{(2\frac{\epsilon_s}{\epsilon_g}\lambda_{g}^{*})^2+\frac{2\epsilon_s\psi_{bi}}{qN}}-2\frac{\epsilon_s}{\epsilon_g}\lambda_{g}^{*}\;.
\end{equation}
It is noted that \eq{eq:FermiUndoped2} and \eq{eq:undoped2} are identical to the counterpart equations of doped graphene (\eq{eq:FermiHighdoped} and \eq{eq:width:doped}) where $\lambda_{g}$ is replaced with $2\lambda_{g}^{*}$. 
\section{Junction capacitance}
\label{sec:capac}
Under external bias voltage ($V$), the width of depletion region and hence the amount of space-charge at the surface of semiconductor would vary. This in turn changes the density of charge induced into graphene which means that the built-in potential of graphene would also vary with the bias voltage. As is depicted in \fig{fig:bandalign}(c), under forward bias the width of depletion region decreases leading to the increment of graphene Fermi level with respect to the equilibrium condition. An opposite situation is occurred under reverse bias (see \fig{fig:bandalign}(d)). The net effect of external bias voltage on the junction properties can be taken into account by substituting $\psi_{bi}$ with $\psi_{bi} - V - kT/q$. The last term ($kT/q$) comes from the majority carrier tail near the edge of depletion layer \cite{Sze_2006}. For the two special limits discussed above (\eq{eq:width:doped} and \eq{eq:undoped2}), the depletion width at graphene-semiconductor contact can be written in the form
\begin{equation}
	\label{eq:width:doped:final}
		w = \sqrt{(\frac{\epsilon_s}{\epsilon_g}\mathcal{L}_{g})^2 + \dfrac{2 \epsilon_s}{q N}(\psi_{bi}-V-\dfrac{kT}{q})} - \frac{\epsilon_s}{\epsilon_g}\mathcal{L}_{g}\;,
\end{equation} 
where $\mathcal{L}_{g}$ is given by
\begin{equation}
	\mathcal{L}_{g}= \begin{cases}
	2\lambda_{g}^{*}\ \ \  &undoped\ graphene\\
	\lambda_{g}     &doped\ graphene
	\end{cases}\;.
\end{equation} 
The differential capacitance may be obtained from $C=dQ_{sc}/dV$, where $Q_{sc}=|qNw|$ is the space-charge per unit area. Utilizing \eq{eq:width:doped:final}, the junction capacitance is obtained as
\begin{equation}
	\label{eq:capacitance}
	C=\left[  (\frac{1}{\epsilon_g}\mathcal{L}_{g})^2+\dfrac{2}{q \epsilon_s N}(\psi_{bi}-V-\dfrac{kT}{q})\right]^{-1/2}\;.
\end{equation}  
It is a common practice to extract junction properties from the slope and intercept of $C^{-2}$ versus $V$ data. At graphene-semiconductor contact we get
\begin{equation}
	\label{eq:c-v}
	C^{-2}= \left\lbrace \left(\dfrac{\mathcal{L}_g}{\epsilon_s}\right)^2+
	\dfrac{2}{q\epsilon_sN}\left(\psi_{bi}-\dfrac{kT}{q}\right) \right\rbrace   
	-\dfrac{2}{q \epsilon_s N}V.
\end{equation}
The main difference between \eq{eq:c-v} and its counterpart equation for the metal-semiconductor junction is the presence of graphene capacitance in the intercept term (curly brackets). Accordingly, the equilibrium capacitance may be written as 
\begin{equation}
\label{eq:cap2}
C^{-2} = C_{g}^{-2} + C_{D}^{-2}\;,
\end{equation}
where $C_{g}=\epsilon_g/\mathcal{L}_{g}$ is the capacitance of graphene and $C_{D}=[q\epsilon_s N/(2(\psi_{bi}-kT/q))]^{1/2}$ is depletion layer capacitance per unit area. We note that $C_g$ becomes increasingly important at the low densities of carrier in graphene-semiconductor system ($n^*<10^{10}\ cm^{-2}$). In the limit of highly-doped graphene, $\mathcal{L}_{g}\rightarrow0$ and the intercept is given by the conventional formula of $2(\psi_{bi}-kT/q)/(q\epsilon_sN)$.

In practice, the barrier heights at graphene-semiconductor junction extracted from C-V measurements are found to be higher than those obtained from current versus temperature (Richardson) measurements \cite{Tongay_2009, Tongay_2012, Shivaraman_2012, Yim_2013, Courtin_2019}. These studies had utilized the conventional metal-semiconductor model to interpret experimental data. Now, it can be verified that there are two reasons for the observed discrepancy. First, in the previous experiments the role of graphene capacitance was not considered and hence total built-in potential was overestimated [\eq{eq:c-v}]. Second, the relation between barrier height and built-in potential at metal-semiconductor junction is given by
\begin{equation}
	\label{eq:C-V:con}
	q\varphi_{Bn}=q\psi_{bi}+q\phi_{n}\;,
\end{equation} 
whereas at graphene-semiconductor interface by combining \eq{eq:g-n:schottky2} and \eq{eq:g-n:builtin} we obtain (see \fig{fig:bandalign}(b))
\begin{equation}
	\label{eq:Sch-Mot}
	q\varphi_{Bn}^{*}=q(\psi_{bi}-\psi_{bi,g})+q\phi_{n}\;,
\end{equation}
which means that the internal potential of graphene was ignored in the calculation of barrier height. In the most practical situations, the contribution of $C_g$ in the junction capacitance is small and the latter effect is responsible for the overestimation of barrier height. The comparison of \eq{eq:C-V:con} with \eq{eq:Sch-Mot} shows that under an ideal condition the difference between the barrier heights extracted from C-V [\eq{eq:C-V:con}] and current-temperature measurements is equal to the built-in potential of graphene. However, the accurate interpretation of C-V data is provided by \eq{eq:c-v} and \eq{eq:Sch-Mot}. 
\section{Barrier height}
\label{sec:SBH}
\subsection{Equilibrium}
\begin{figure}[]
	\includegraphics[scale=1.0]{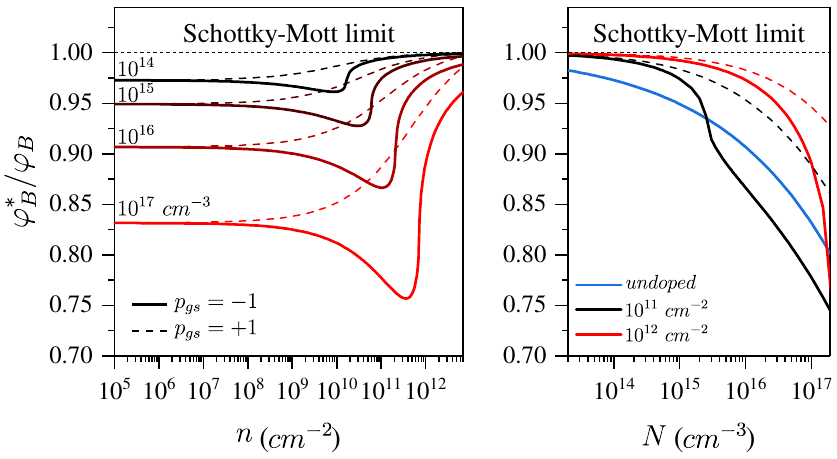}
	\caption{
		\label{fig:SBH} 
		Equilibrium barrier height ($\varphi_{B}^{*}/\varphi_{B}$) versus (left) the density of charge carrier in isolated graphene and (right) the concentration of dopants in semiconductor. 
	}
\end{figure}
Following the discussions given in sections \ref{sec:energy} and \ref{sec:capac}, the barrier height at graphene-semiconductor contact may be written in a general form
\begin{equation}
	\label{eq:SBHg}
	q\varphi_{B}^{*}(V)=q\varphi_{B}-q\Delta\varphi_{B}(V)\;,	
\end{equation}
where $q\varphi_{B}$ is barrier height in the Schottky-Mott limit. Neglecting image-force effect, $q\Delta\varphi_{B}(V)\!=\!q\Delta\phi_{g}(V)$ which is given by \eq{eq:FermiGeneral}. \fig{fig:SBH} represents equilibrium (zero-bias) barrier height at graphene-silicon interface calculated through numerical solution of \eq{eq:width}. The barrier height at graphene-semiconductor contact is generally smaller than the Schottky-Mott limit. The deviation increases with $(i)$ decreasing the density of charge carrier in isolated graphene, and $(ii)$ increasing the concentration of dopants in semiconductor. Only in the limit of highly-doped graphene ($n\!\sim\!\!10^{13}\ cm^{-2}$ or $E_{F,g}\!>\!\!300\ meV$), the barrier height approaches to the Schottky-Mott limit. In addition, the magnitude of deviation for junctions with diverse polarity of graphene and semiconductor $p_{gs}=-1$ is higher than the junctions with identical polarities $p_{gs}=+1$ (see section \ref{sec:builtin}). Also the deviation from Schottky-Mott limit for $p_{gs}=-1$ could be higher than the undoped graphene due to the polarity flip effect (see right panel of \fig{fig:SBH}). In practice, CVD-grown graphene layers often display p-type character ($p_g=+1$) with Fermi level ranging between $E_{F,g}\approx50-200\ meV$ ($n\approx2\times10^{11}-2\times10^{12}\ cm^{-2}$). Hence, the deviation of realistic graphene$-$p-Si samples from Schottky-Mott limit is less than the graphene$-$n-Si samples. Although the practical deviation is around 0.85 - 0.95, it has a high impact on the reverse saturation current.
\subsection{Bias-driven tunable barrier height}
An important consequence of graphene Fermi level variation with external bias (section \ref{sec:capac}) is that the height of barrier varies with bias voltage which is known as ``tunable barrier height'' in the literature \cite{Yang_2012, Tongay_2012, An_2013, Di_Bartolomeo_2016}. In effect, the height of barrier decreases with reverse bias voltage and vice versa. Since charge transport depends exponentially on $q\varphi_{B}^{*}$, the reduction of barrier height leads to noticeable increment of saturation current with reverse bias voltage. From the above discussions it is clear that the tunable barrier height stems from the internal potential of graphene. For the two special limits discussed in section \ref{sec:twolim}, by combining Eqs. (\ref{eq:FermiHighdoped}), (\ref{eq:width:doped:final}), and (\ref{eq:SBHg}) the tunable barrier height is obtained as
\begin{eqnarray}
	\label{eq:SBH:bias}
	&&q\varphi_{B}^{*}=q\varphi_{B}-\frac{q^2}{\epsilon_g}\mathcal{L}_{g}N\times \nonumber\\
	&&\left\lbrace \sqrt{(\frac{\epsilon_s}{\epsilon_g}\mathcal{L}_{g})^2 + \dfrac{2 \epsilon_s}{q N}(\psi_{bi}-V-\dfrac{kT}{q})} - \frac{\epsilon_s}{\epsilon_g}\mathcal{L}_{g} \right\rbrace\;. 
\end{eqnarray}
There is a close relation between barrier height and junction capacitance. Differentiating \eq{eq:SBH:bias} we get
\begin{eqnarray}
	\label{eq:dSBHdv}
	\frac{d}{dV}&&(q\varphi_{B}^{*})=\frac{q}{\epsilon_g}\mathcal{L}_{g}\times \nonumber \\ &&\left[ (\frac{1}{\epsilon_g}\mathcal{L}_{g})^2 + \dfrac{2 }{q \epsilon_s N}(\psi_{bi}-V-\dfrac{kT}{q})\right] ^{-1/2}\;.
\end{eqnarray}
The term inside square brackets is just junction capacitance [\eq{eq:capacitance}] and the prefactor is the capacitance of graphene [\eq{eq:cap2}]. It is more instructive to write the foregoing result in the form 
\begin{equation}
	\label{eq:dSBHdv2}
	\frac{d\varphi_{B}^{*}}{dV}=\frac{C}{C_g}\;.
\end{equation}
Hence, the derivative of barrier height with respect to voltage is directly proportional to the junction capacitance and proportionality coefficient is given by the inverse of graphene capacitance. \eq{eq:dSBHdv2} provides a direct method to determine the capacitance (and hence, the density of charge carrier, Fermi level, and built-in potential) of graphene in the graphene-semiconductor system. Experimentally, $q\varphi_{B}^{*}(V)$ and $C(V)$ can be extracted from temperature dependency of current-voltage characteristics and capacitance-voltage measurements, respectively.
\subsection{The impact of image-force}
\begin{figure}[]
	\includegraphics[scale=1.0]{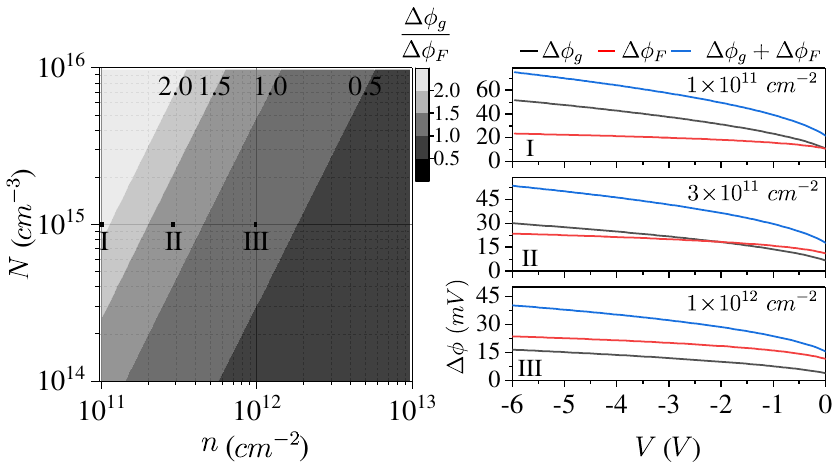}
	\caption{
		\label{fig:img} 
		Left: Relative magnitude of $\Delta\phi_{g}$ with respect to $\Delta\phi_{F}$ at doped graphene-silicon contact ($V=-5\ V$). Right: Built-in potential of graphene, image-force, and total barrier lowering versus bias voltage at the indicated regions in the left panel.
	}
\end{figure}
The image-force (or Schottky) effect is the image-force-induced lowering of barrier height at metal-semiconductor junction. The image-force lowering also affects the barrier height at graphene-semiconductor interface. In order to assess the impact of image-force effect, we may consider the bias-dependent term of barrier height as (see \eq{eq:SBHg})
\begin{equation}
	\label{eq:SBH:imf}
	q\Delta\varphi_{B}(V) = q(\Delta\phi_{g} + \Delta\phi_{F})\;,
\end{equation}
where $q\Delta\phi_{g}$ is the built-in potential of graphene (barrier tunability effect) given by the second term in the RHS of \eq{eq:SBH:bias}, and $q\Delta\phi_{F}$ is the image-force lowering effect given by
\begin{equation}
	\label{eq:imf}
	\Delta\phi_{F} = \left[ \frac{q^3 N \psi_{bi,s}}{8\pi^2\epsilon_s^3} \right]^{1/4}\;,
\end{equation}
with
\begin{equation}
	\label{eq:bips}
	\psi_{bi,s} = \psi_{bi}-V-\dfrac{kT}{q} - \psi_{bi,g}\;.
\end{equation} 
In the limit of undoped graphene, by combining Eqs. (\ref{eq:SBuilt-in}), (\ref{eq:FermiUndoped}), and (\ref{eq:imf}) we get
\begin{equation}
	\label{eq:imf:undoped}
	\frac{\Delta\phi_{g}}{\Delta\phi_{F}}=\frac{2\pi\epsilon_s\hbar v_F}{q^2}=2.63\;.
\end{equation}
Hence, the voltage dependency of barrier height at undoped graphene-semiconductor contact is modulated by the graphene internal potential. For the other limit of doped graphene, the relative impact of image-force with respect to the built-in potential of graphene is very sensitive to $N$, $n$, and bias voltage. \fig{fig:img} represents the relative magnitude of $\Delta\phi_{g}$ with respect to $\Delta\phi_{F}$. In the regime of doped graphene, $\Delta\phi_{g}/\Delta\phi_{F}$ increases with the dopant concentration in semiconductor because $\Delta\phi_{g} \propto N^{1/2}$ while $\Delta\phi_{F} \propto N^{1/4}$. 
At a fixed dopant concentration, $\Delta\phi_{g}/\Delta\phi_{F}$ decreases with the density of charge carriers in the isolated graphene since $\Delta\phi_{g} \propto n^{-1}$. For example, consider constant dopant concentration of $N=10^{15}\ cm^{-3}$ in \fig{fig:img}. At $n\lesssim10^{11}\ cm^{-2}$, $\Delta\phi_{g}$ is the dominant effect in the reduction of barrier height (region I) whereas for $n\gtrsim10^{12}\ cm^{-2}$ the barrier reduction is modulated by $\Delta\phi_{F}$ (region III). In addition, since $\Delta\phi_{g} \propto V^{1/2}$ and $\Delta\phi_{F} \propto V^{1/4}$, at the intermediate densities ($10^{11}<n<10^{12}\ cm^{-2}$), there is a crossover from $\Delta\phi_{F}$ at low reverse voltages to $\Delta\phi_{g}$ at high reverse voltages (region II). 
\section{Inversion}
\label{sec:inv}
Considering the work function of graphene at band-crossing point ($q\phi_{g}^o=4.6\ eV$) \cite{Yu_2009} and the fact that practical (CVD-grown) graphene samples are often unintentionally p-doped with $E_{F,g}\approx 50-200\ eV$, the contact formation between graphene and n-type silicon may lead to severe band bending at the silicon surface and result in inversion. Upon inversion, the density of holes at the surface of n-type silicon becomes higher than density of excess electrons at the surface (weak inversion) or even in the bulk of n-Si (strong inversion). The Formation of inversion layer induces an internal p-n junction at the surface of silicon and is of great importance for applied purposes. By introducing Fermi potential energy of semiconductor as $q\Psi_{B} = E_{F,s} - E_{i}$ (note that q$\Psi_{B}$ is defined with respect to the intrinsic Fermi level $E_i$ in the bulk of semiconductor, see \fig{fig:bandalign}(a) and \fig{fig:bandalign}(b)), various phases of silicon surface at graphene$-$n-Si contact may be distinguished as: depletion if $0<\psi_{bi,s}/\Psi_{B}<1$, weak inversion if $1\leq\psi_{bi,s}/\Psi_{B}<2$, and strong inversion if $2\leq\psi_{bi,s}/\Psi_{B}$.

\begin{figure}[]
	\includegraphics[scale=1.0]{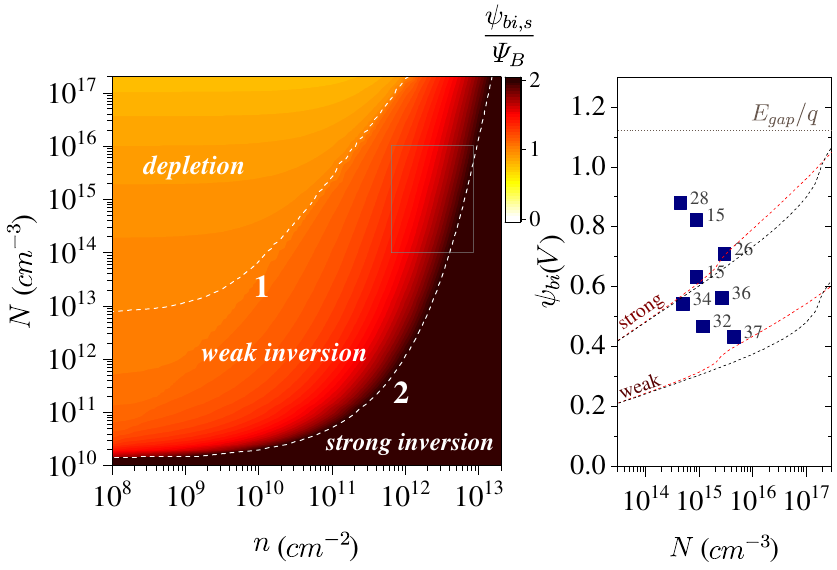}
	\caption{
		\label{fig:inv} 
		Left: Different phases of silicon surface at p-doped graphene$-$n-type silicon contact. Right: Experimental data of (total) built-in potential versus dopant concentration obtained in various experimental studies on graphene$-$n-Si. The corresponding references are shown beside the data. The red and black dash-lines indicate the onset of weak (and strong) inversion at $n=10^{11}$ and $10^{12}\ cm^{-2}$, respectively.
	}
\end{figure}

The relative magnitude of built-in potential at silicon surface with respect to the Fermi potential is presented in the left panel of \fig{fig:inv}. The data of this plot is obtained by numerical solution of \eq{eq:width} for p-doped graphene$-$n-type silicon system. Depletion, weak inversion, and strong inversion regions are indicated in the figure. The surface tends to strong inversion with ($i$) increasing the density of charge carrier in graphene and ($ii$) decreasing the concentration of donors in silicon. The rectangular window in this figure approximately indicates the zone of practical samples which locates at the periphery of weak and strong inversion. Right panel of \fig{fig:inv} represents experimental data of total built-in potential versus donor concentration of graphene$-$n-Si samples reported by various groups \cite{Miao_2012, Tongay_2012, Tongay_2009, Di_Bartolomeo_2017, Yim_2013, Zhong_2016, Yu_2016}. Considering the onset of weak and strong inversion, we conclude that in all of the practical graphene$-$n-Si samples the surface of silicon is inverted and almost in half of the cases a strong inversion layer is developed at the surface of silicon.
\section{Conclusion}
\label{sec:conclu}
The standard model of semiconductor junctions is developed to graphene-semiconductor contact. Due to the finite density of states, a fraction of total built-in potential is dropped across graphene layer. Nonetheless, the depletion region is constructed entirely on the surface of semiconductor by the virtue of graphene atomic thickness. In addition, the Schottky-Mott limit and the sum rule of barrier heights are violated at graphene-semiconductor contact. Considering the above elements, it is instructive to refer this contact as semimetal-semiconductor junction in order to distinguish it from the Schottky junction. The internal potential of graphene serves as a criterion to recognize semimetal-semiconductor and metal-semiconductor junctions. The graphene-semiconductor contact approaches to a true Schottky junction only in the limit of highly-doped graphene. \\
Doping polarity of graphene and semiconductor ($p_{gs}=p_gp_s$) play a crucial role in the built-in potential of graphene. Diverse polarities ($p_{gs}=-1$) give rise to higher built-in potentials. The width of depletion region and the junction total capacitance are correlated, respectively, with the Thomas-Fermi length of carriers and the capacitance of graphene. The origin of large barrier heights extracted form C-V measurements is demonstrated. In an ideal condition, the difference between the barrier heights extracted from C-V and current-temperature measurements is equal to the graphene internal potential. Also it is shown that the voltage derivative of tunable barrier is directly (inversely) proportional to the junction (graphene) capacitance. This provides a straightforward method to determine the electronic properties of graphene in the graphene-semiconductor system.\\
Additionally, the relative impact of image-force effect at the graphene-silicon interface is studied. For undoped graphene, the barrier lowering effect is modulated by the graphene internal potential. In the limit of doped graphene, however, the dominant mechanism is mainly determined the density of charge carriers in graphene and bias voltage. Finally, various surface phases at the graphene-silicon interface are distinguished. A comparison with experimental data uncovers the presence of strong inversion layer at the surface of silicon.

\begin{acknowledgments}
Financial support from ``Iran's National Elites Foundation'' is acknowledged.
\end{acknowledgments}
\appendix
\renewcommand\thefigure{\thesection \arabic{figure}}
\renewcommand\thetable{\thesection \arabic{table}} 

\section{Work function of semiconductor }
\label{appendixA}
\setcounter{figure}{0}
The work function of a n-type semiconductor is given by
\begin{equation}
	\label{eq:SworkfunctionPs-1}
	q\phi_s = q(\chi + \phi_n) = q\chi + (E_c - E_{F,s})\;.
\end{equation}
Assuming a complete ionization, the density of free electrons ($n_{no}$) can be approximated by the concentration of dopants ($N$) as 
\begin{equation}
	\label{eq:FCn}
	n_{no} = N_c\ exp(-(E_c - E_{F,s})/kT) \cong N\;,
\end{equation}
where $N_c$ in the effective density of states at the edge of conduction band. Rearranging \eq{eq:FCn}, one gets
\begin{equation}
	\label{eq:phi_n}
	q\phi_n = E_c - E_{F,s} \cong (kT)Ln(N_c/N)\;.
\end{equation}
Similarly, for a p-type semiconductor we have
\begin{eqnarray}	
\label{eq:phi_p}
q\phi_s\ &&= q\chi + E_{gap} - q\phi_p\;,\\
q\phi_p\ &&= E_{F,s} - E_v \;, \\
p_{po}\ &&= N_v\ exp(-(E_{F,s}-E_v)/kT) \cong N \;,\\
q\phi_p\ &&\cong (kT)Ln(N_v/N)\;,
\end{eqnarray}
where $p_{po}$, $N$, and $N_{v}$ are the density of free holes, the concentration of acceptors, and the effective density of states at the edge of valance band.
\section{The limit of doped graphene}
\label{appendixB}
\setcounter{figure}{0}
The doped graphene-semiconductor contact is characterized by $n\gg Nw$. \fig{fig:adj} shows the magnitude of $n/(Nw)$ calculated by numerical solution of \eq{eq:width}. To a good approximation, the limit of doped graphene is valid for $N<10^{16}\ cm^{-3}$ and $n>10^{12}\ cm^{-2}$. 
\begin{figure}[hhh]
	\includegraphics[scale=0.8]{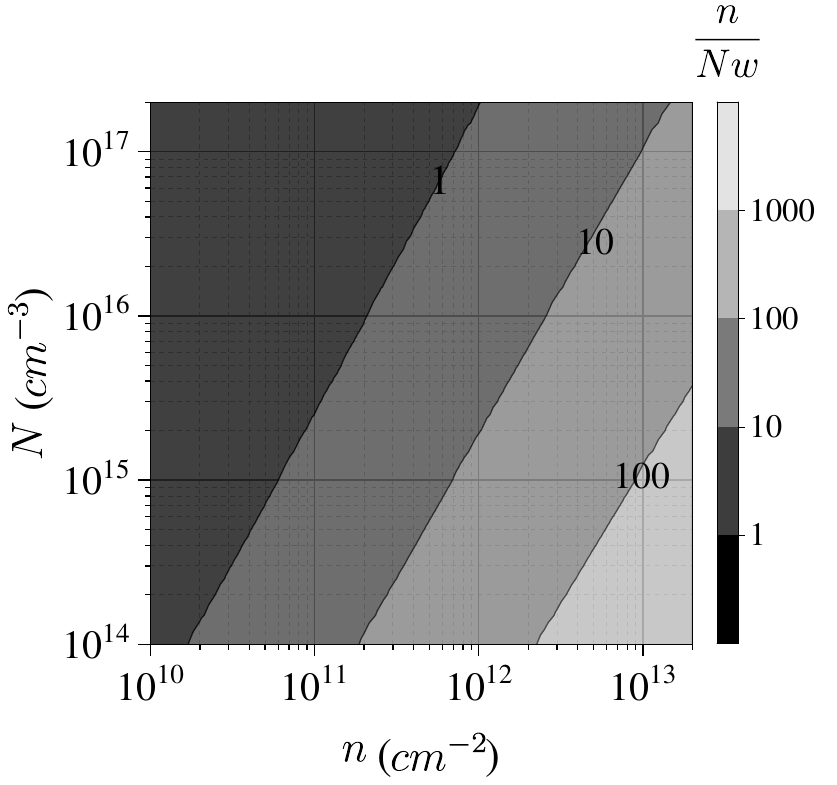}
	\caption{
		\label{fig:adj} 
		The applicability domain of doped-graphene limit. Major levels are indicated in the graph.
	}
\end{figure}
\section{Electronic properties}
\label{appendixX}
\setcounter{figure}{0}
\tab{tab:constants} represents the electronic properties of graphene and silicon used in the numerical calculations.
\begin{table}[hhb]
	\caption{
		\label{tab:constants}
		 Electronic properties of graphene and silicon.
	}
\begin{ruledtabular}
	\begin{tabular}{lccc}
		\textrm{Quantity}&
		\textrm{Symbol}&
		\textrm{Silicon}&
		\textrm{Graphene}\\
		\colrule
		Dielectric constant & $\epsilon$ & 12 & 5\\
		Work function (V) & $\phi_{g}^{o}$ & $-$ & 4.60\\
		Fermi velocity ($cm/s$) & $v_F$ & $-$ & 10$^{8}$\\
		Electron affinity ($V$) & $\chi$ & 4.05 & $-$\\
		Band gap energy ($eV$) & $E_{gap}$ & 1.12 & $-$\\
		Effective DOS in CB ($cm^{-3}$) & $N_c$ & 2.80$\times$10$^{19}$ & $-$\\
		Effective DOS in VB ($cm^{-3}$) & $N_v$ & 2.65$\times$10$^{19}$ & $-$\\
	\end{tabular}
\end{ruledtabular}
\end{table}
\bibliography{REFMSGSJ}
\end{document}